\documentclass[aps,prd,onecolumn,notitlepage,preprint,showpacs, showkeys,superscriptaddress]{revtex4-2}

\usepackage{graphicx}
\usepackage{epstopdf}
\usepackage{amsmath}
\usepackage[section]{placeins}
\usepackage{slashed}
\usepackage{color}
\usepackage{soul}
\usepackage{appendix}
\usepackage{placeins}
\usepackage[utf8]{inputenc}
\usepackage{pstricks}
\usepackage{multirow}
\usepackage{hyperref}
\usepackage{subfigure}
\usepackage{epsfig}
\usepackage{cancel}
\usepackage{braket}
\usepackage[normalem]{ulem} 
\usepackage{longtable,booktabs}
\newcommand{\be}{\begin{equation}}
\newcommand{\ee}{\end{equation}}
\newcommand{\ben}{\begin{eqnarray}}
\newcommand{\een}{\end{eqnarray}}

\begin{document}

\title{Nature of the newly found $\Omega(2109)$}

\author{Ta\'isa Veloso}
\email[]{taisa.veloso@unifesp.br}
\affiliation{Universidade Federal de S\~ao Paulo, C.P. 01302-907, S\~ao Paulo, Brazil}

\author{K.~P.~Khemchandani}
\email[]{kanchan.khemchandani@unifesp.br}
\affiliation{Universidade Federal de S\~ao Paulo, C.P. 01302-907, S\~ao Paulo, Brazil}

\author{A.~Mart\'inez~Torres}
\email[]{amartine@if.usp.br}
\affiliation{Instituto de F\'{\i}sica, Universidade de S\~{a}o Paulo, Rua do Mat\~{a}o, São Paulo SP, 05508-090, Brazil}   

\author{H. Nagahiro}
\email[]{nagahiro@rcnp.osaka-u.ac.jp}
\affiliation{Department of Physics, Nara Women’s University, Nara 630-8506, Japan}   
\affiliation{Research Center for Nuclear Physics, Osaka University, Ibaraki, Osaka 567-0047, Japan}

\author{A.~Hosaka}
\email[]{hosaka@rcnp.osaka-u.ac.jp}
\affiliation{Research Center for Nuclear Physics, Osaka University, Ibaraki, Osaka 567-0047, Japan}

\vspace{1cm}

\begin{abstract}
We present model calculations to reveal the nature of the newly found $\Omega^-(2109)$ by the BESIII Collaboration, and show that the state has a strong correlation with the $\bar K^*(892)\Xi$ system. Our study is based on solving scattering equations in a coupled channel approach, which involves $K^-\Xi^0$, $\bar K^0\Xi^-$, $K^{*-}\Xi^0$, and $\bar K^{*0}\Xi^-$. We obtain the lowest order scattering amplitudes for different spin and isospin cases and find that an isoscalar state with spin-parity $1/2^-$ is generated with precisely the same mass as $\Omega^-(2109)$. We do not find any state with total spin 3/2, nor do we find any state in the isovector sector. We determine correlation functions to encourage such an experimental study and confirm the nature of $\Omega^-(2109)$.
\end{abstract}

\maketitle
\newpage
\section{\label{sec:level1} Introduction}
The present work is motivated by the recent study of the process $e^+ e^-\to \Omega^{*-}\bar \Omega^++ c.c$ by the BESIII collaboration~\cite{BESIII:2024eqk} investigated in the range of center-of-mass energies varying from 4130 MeV to 4700 MeV.  The former work reports the first evidence for the existence of a new excited $\Omega$ state with mass around 2100 MeV, denominated in Ref.~\cite{BESIII:2024eqk} as $\Omega^-(2109)$, besides the confirmation of the existing $\Omega^-(2012)$~\cite{PDG2024}. The precise findings on the mass, $M$, and width, $\Gamma$, of $\Omega^-(2109)$ obtained through a Breit-Wigner fit performed in Ref.~\cite{BESIII:2024eqk}, keeping the values for $\Omega^-(2012)$ fixed as catalogued by the Particle Data Group (PDG)~\cite{PDG2024},  are as stated below:
\begin{align}\label{OM2109}
M &=2108.5\pm 5.2\pm0.9 \text{ MeV},\\\nonumber
\Gamma &=18.3\pm16.4\pm 5.7 \text{ MeV}.
\end{align}
The spin-parity for $\Omega^-(2109)$  is assumed to be $J^P=3/2^-$ in Ref.~\cite{BESIII:2024eqk} inspired by the lattice QCD results obtained by the Hadron Spectrum Collaboration~\cite{Edwards:2012fx} and it is inferred in Ref.~\cite{BESIII:2024eqk} that both $\Omega^-(2012)$ and $\Omega^-(2109)$ could be conventional baryons. However, it should be pointed out that different quark model calculations~\cite{Isgur:1978xj,Crede:2013kia,Xiao:2018pwe} seem to predict the existence of an almost spin degenerate pair in the mass region $1950-2020$ MeV which is lower than 2109 MeV. A recent quark model~\cite{Luo:2025cqs} study also seems to fail in finding a mass compatible with the state found by the BESIII Collaboration~\cite{BESIII:2024eqk}.
Furthermore, a QCD sum rule analysis predicts a doublet of spin-parity $1/2^-$ and $3/2^-$ near the nominal mass  region of $\Omega(2012)$~\cite{Su:2024lzy}. The $1/2^-\left(3/2^-\right)$ state found in the former work has properties compatible with $\Omega(2109)~\left[\Omega(2012)\right]$ when considering error bars. 
In case of lattice QCD~\cite{Edwards:2012fx} as well, two states have also been found, though with mass values that are distant from each other. In any case, it is important to recall that a direct comparison with the results from lattice QCD should be done with some care, as the simulations in Ref.~\cite{Edwards:2012fx} are carried out using a pion mass of 391 MeV. Besides, extrapolating the results to an infinite volume can lead to some differences in the mass values. 

An interesting experimental finding to be noted in this context is that data on the $\bar K\Xi$ invariant mass are available from BELLE~\cite{Belle:2018mqs}, and more recently from the ALICE  collaboration~\cite{ALICE:2025atb}, which do not show any significant signal around 2100 MeV, although the presence of $\Omega^-(2012)$ is clearly seen. It is then important to reconcile the results of Refs.~\cite{ALICE:2025atb,Belle:2018mqs,BESIII:2024eqk}.  

To test the possibility of relating the newly found excited $\Omega$-state with meson-baryon dynamics, we solve the Bethe-Salpeter equation in a coupled channel approach by considering the $K^-\Xi^0$, $\bar K^0\Xi^-$, $K^{*-}\Xi^0$ and $\bar K^{*0}\Xi^-$ systems. The formalism relies on our previous works~\cite{Khemchandani:2011et,Khemchandani:2011mf}, where pseudoscalar-baryon interactions are determined from the lowest order chiral lagrangian~\cite{Gasser:1983yg} and vector-baryon interactions are obtained from hidden local symmetry~\cite{Bando:1985rf}, which is based on the treatment of vector mesons as gauge bosons of the theory. The transition between pseudoscalar-baryon and vector-baryon channels is evaluated through a contact term coming from the consideration of the minimal coupling when treating vector mesons as gauge bosons again. To determine the kernels required to solve the scattering equation, we calculate the $t$- and $u$-channel diagrams, and a contact term is also considered in the case of vector-baryon systems. To summarize our results, we find that an isoscalar state with spin-parity $1/2^-$ gets dynamically generated, whose properties are in very good agreement with those summarized in Eq.~(\ref{OM2109}). Our state can be interpreted as a $\bar K^*\Xi$ quasibound state, which has a small coupling to the pseudoscalar-baryon systems. The findings of this work can be useful to explain the missing signal of $\Omega^-(2109)$ in the data of Refs.~\cite{ALICE:2025atb, Belle:2018mqs}. It should be added that no isoscalar states are found in the spin 3/2 configuration, and none are found in the isospin 1 case for both spins. With the purpose of motivating further experimental searches, we calculate femtoscopic correlation functions for both pseudoscalar-baryon and vector-baryon channels. 

In the following, we first present details of the mathematical framework used, and, subsequently, we discuss the results of our work. 

\section{Theoretical framework}
\begin{figure}[h!]
    \centering
    \includegraphics[width=\textwidth]{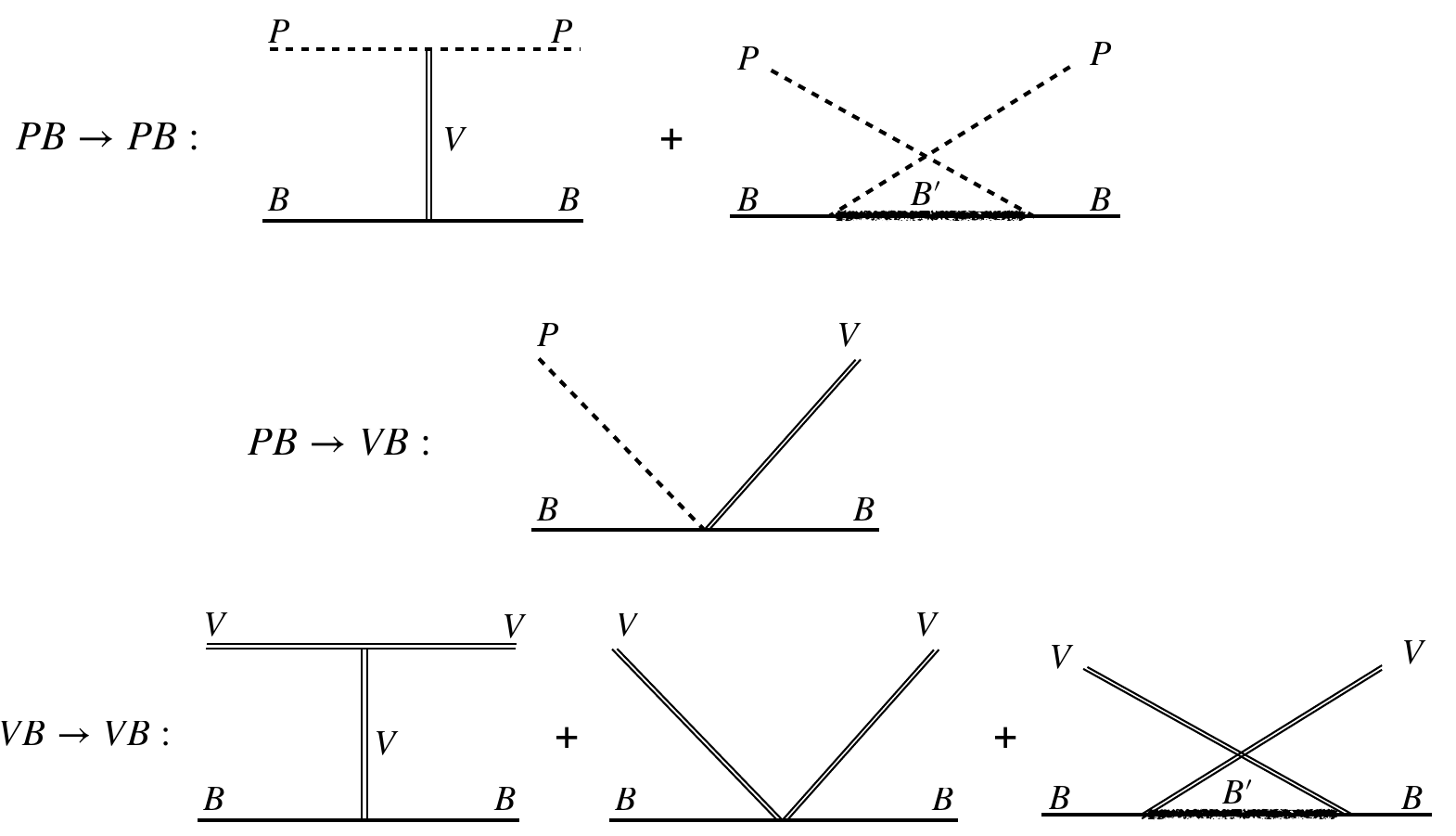}
    \caption{Different diagrams contributing to the lowest order amplitudes, $\mathcal{V}$. Double (dashed) lines in these diagrams represent vector (pseudoscalar) mesons, while the baryons are shown as thick lines. Smeared lines stand for baryons exchanged in the $u$-channel.}
    \label{diags}
\end{figure}

The present work studies meson-baryon scattering in a coupled channel approach. We consider pseudoscalar and vector mesons interacting with octet baryons. The idea is to deduce the $T$-matrix by solving the Bethe-Salpeter equation 
\begin{align}\label{bseq}
\mathcal{T}=\mathcal{V}+\mathcal{V}G\mathcal{T},
\end{align}
in its on-shell factorized form, which relies on (1) the reabsorption of the off-shell dependence of the amplitudes into the renormalization of masses and pion decay constant to their physical values and (2) a suitable regularization of the two-body loop function  (see Ref.~\cite{Khemchandani:2018amu}).
The lowest order scattering amplitudes required to solve Eq.~(\ref{bseq}) are determined from the Lagrangians given in  Refs.~\cite{Khemchandani:2011et,Khemchandani:2011mf,Khemchandani:2018amu}.  We find it instructive to summarize the different Lagrangians needed to compute the diagrams shown in Fig.~\ref{diags} that sum up to give the kernels for Eq.~(\ref{bseq}). 

The Lagrangian that describes vector-baryon interactions is written, following Ref.~\cite{Khemchandani:2011et}, as 
\begin{eqnarray} \label{vbb}
\mathcal{L}_{VB}&=& -g_{VB} \Biggl\{ \langle \bar{B} \gamma_\mu \left[ V^\mu, B \right] \rangle + \langle \bar{B} \gamma_\mu B \rangle  \langle  V^\mu \rangle  
\Biggr. \\ \nonumber
&+&\left. \frac{1}{4 M} \left( F \langle \bar{B} \sigma_{\mu\nu} \left[ V^{\mu\nu}, B \right] \rangle  + D \langle \bar{B} \sigma_{\mu\nu} \left\{ V^{\mu\nu}, B \right\} \rangle\right)\right\}.
\end{eqnarray}
The formulation of this Lagrangian in Ref.~\cite{Khemchandani:2011et} was motivated by the idea of hidden local symmetry, in which vector mesons play the role of gauge bosons.~\cite{Bando:1985rf}. The meson and baryon fields in Eq.~(\ref{vbb}) represent the corresponding SU(3) matrices, and the tensor field $V^{\mu\nu}$ is written, in our convention, as 
\begin{equation}
V^{\mu\nu} = \partial^{\mu} V^\nu - \partial^{\nu} V^\mu + ig \left[V^\mu, V^\nu \right]. \label{tensor}
\end{equation}
The values of the constants $D$ = 2.4 and $F$ = 0.82 in Eq.~(\ref{vbb}) are set to reproduce the magnetic moments of the baryons~\cite{Jido:2002yz}. 

The pseudoscalar-baryon interactions are determined using the standard lowest order chiral Lagrangian~\cite{Weinberg:1968de,Bernard:1995dp,Ecker:1994gg,Ramos:2002xh}. Finally, the transition amplitudes between the vector-baryon and pseudoscalar-baryon channels are determined through
\begin{align}
\mathcal{L}_{PBVB} = \frac{-i g_{KR}}{2 f_\pi} \left ( F^\prime \langle \bar{B} \gamma_\mu \gamma_5 \left[ \left[ P, V_\mu \right], B \right] \rangle + 
D^\prime \langle \bar{B} \gamma_\mu \gamma_5 \left\{ \left[ P, V_\mu \right], B \right\}  \rangle \right), \label{pbvb}
\end{align}
which is an extension of the Kroll-Rudermann term that describes the photon-induced pion production. The former interaction is obtained by introducing the photon through the minimal coupling procedure. To obtain the Lagrangian in Eq.(\ref{pbvb}) from the Kroll-Rudermann one, inspired by the hidden local symmetry and the  vector meson dominance embedded in this approach, we introduce vector mesons as the gauge bosons, associated with the hidden local symmetry~\cite{Bando:1985rf}. In addition, we consider the value of $g_{KR}$ to be $m_\rho/(\sqrt{2}f_\pi)\sim 6$ in accordance with the Kawarabayashi-Suzuki-Fayyazuddin-Riazuddin (KSFR) relation~\cite{Kawarabayashi:1966kd,Riazuddin:1966sw}. The constants $F^\prime = 0.46$ and $D^\prime=0.8$ in Eq.~(\ref{pbvb}) are in line with empirically known values~\cite{Yamanishi:2007zza} and lead to a value of the axial coupling of the nucleon,  $F + D \simeq  g_A = 1.26$, in agreement with the experimental data.

We now summarize the lowest order amplitudes for the different diagrams shown in Fig.~\ref{diags}, as determined from the aforementioned Lagrangians, after projecting them on spin 1/2 and on angular momentum $L=0$ (S-wave), and expressing them in the center-of-mass frame.
\begin{itemize}
    \item The t-channel exchange for pseudoscalar-baryon (PB) and vector-baryon (VB), has the same aspect after spin projection~\cite{Oset:2010tof,Khemchandani:2018amu,Khemchandani:2011et}. We have   
     \begin{align}\label{eq:t}
              \mathcal{V}_{t,\,\text{MB}}\left(i\to j\right)=\frac{1}{4f_\pi^2}\left(2\sqrt{s}-E_i-E_j\right) \sqrt{\frac{(E_i+M_i)}{2M_i} \frac{(E_j+M_j)}{2M_j}},
        \end{align}
        where MB in the subscript indicates that the given amplitude is  for any of the meson-baryon systems considered in this work.
     In Eq.~(\ref{eq:t}), and throughout the manuscript, $E_i \left(E_j\right)$ and  $M_i \left(M_j\right)$, respectively, stand for the energy and mass of the incoming (outgoing) baryon and $f_\pi = 93 \text{ MeV}$.
    It must be emphasized that this amplitude is spin independent, in case of  VB channels, and is  the same for all channels (except for the mass dependence).

    \item For the vector-baryon u-channel exchange diagram, with total spin $J=\frac{1}{2}$, in the nonrelativistic approximation, we obtain 
    \begin{align}
    \label{uvb}
\mathcal{V}_{u,\text{VB}}^{1/2}\left(i\to j\right)=-U_{ij,\text{VB}} ~\frac{g_i g_j}{2\tilde{M}-m_v}\sqrt{\frac{(E_i+M_i)}{2M_i} \frac{(E_j+M_j)}{2M_j}}, 
    \end{align}
  where the $U_{ij,\text{VB}}$ coefficients for different processes are listed in Table~\ref{uij} of the Appendix. Further, $\tilde{M}$ represents the average mass of the exchanged baryon and $m_v$ is the isospin-average mass of $\bar K^*(892)$, relevant for this study. The coupling $g_i~(g_j) = \frac{m_i(m_j)}{\sqrt{2}f_\pi}$, as mentioned before, is defined through the KSRF relation, with $m_i (m_j)$ being the mass of the incoming (outgoing) meson. 
  
  It is useful to mention here that a fully relativistic treatment of vector-baryon amplitudes was investigated in Ref.~\cite{Khemchandani:2016ftn}, and it was found that a nonrelativistic approach led to  similar results. Hence, we have kept the VB amplitudes as determined within a nonrelativistic approximation.
 
   For the pseudoscalar-baryon processes, we determine relativistic amplitudes, keeping in mind the smaller masses of pseudoscalar mesons. In this way, we write the PB amplitude, proceeding through a $u$-channel exchange of an octet baryon, after projecting on the S-wave as (following Refs.~\cite{Khemchandani:2018amu,Oller:2000fj}),
   \begin{align}
        \mathcal{V}_{u,\text{PB}}\left(i\to j\right)&=\sum_k\left(-\frac{U_{ij,\text{PB}}^{k}}{4f_\pi^2} \right)\sqrt{\frac{(E_i+M_i)}{2M_i} \frac{(E_j+M_j)}{2M_j}} \Biggl\{ \sqrt{s}+M_k- (M_i+M_k)(M_j+M_k) \nonumber \\&\times \frac{(\sqrt{s}+M_i+M_j-M_k)}{2(E_i+M_i)(E_j+M_j)} + \frac{(M_i+M_k)(M_j+M_k)}{4|\,\vec{p_i}\,||\,\vec{p_j}\,| } \Bigg(\sqrt{s}-M_i-M_j+M_k\nonumber \\   
        &-\bigl(s+M_k^2-\left(m_i^2+m_j^{2} + 2E_iE_j\right)\bigr) \frac{(\sqrt{s}+M_i+M_j-M_k)}{2(E_i+M_i)(E_j+M_j)} \Bigg)\nonumber \\  
        &\times   \ln \bigg[\frac{s+M_k^2-(m_i^2+m_j^{2} +2E_iE_j)-2|\,\vec{p_i}\,||\,\vec{p_j}\,| }{s+M_k^2-(m_i^2+m_j^{2} +2E_iE_j)+2|\,\vec{p_i}\,||\,\vec{p_j}\,| } \bigg] \Biggr\},  \label{uchnpb} 
    \end{align}
which, at the threshold (i.e., when $|\vec{p_i}| = 0$ and/or $|\vec{p_j}| = 0$), gets simplified to
    \begin{align}
        \mathcal{V}_{u,\text{PB}} (i\to j) &= \sum_k\left(-\frac{U_{ij,\text{PB}}^{k}}{4f_\pi^2}\right)\sqrt{\frac{(E_i+M_i)}{2M_i} \frac{(E_j+M_j)}{2M_j}}\frac{1}{u-M_k^2}\nonumber  \\  &\times\Biggl\{ u\big[ \sqrt{s} +M_k \big]  + \sqrt{s} \big[M_j (M_i+M_k)+M_iM_k \big]\nonumber \\  
        &- \big[M_j(M_i+M_k)(M_i+M_j)  + M_i^2M_k\big] \Biggr\}.  \label{upb}
    \end{align}  
  The $U_{ij,\text{PB}}^{k}$ coefficients in Eqs.~(\ref{uchnpb}) and~(\ref{upb}) are given in Table~\ref{uijk}, $|\,\vec{p_i}\,|~ \left(|\,\vec{p_j}\,|\right)$ represent the incoming (outgoing) center-of-mass momentum, and $M_k$ is the mass of the baryon being exchanged.
    It is also relevant to mention that a monopole form factor is multiplied to the $u$-channel amplitude, to account for the suppression of baryon-antibaryon production (annihilation). 

    \item A contact-term (CT), as shown in Fig.~\ref{diags} for $VB$ processes, arises from the commutator part of the $V^{\mu\nu}$ tensor when used in Eq.~(\ref{vbb}). The corresponding $J=\frac{1}{2}$ contribution is
    \begin{align}
    \label{ct}
      \mathcal{V}_{CT,\text{VB}}^{1/2}\left(i\to j\right)=C_{ij}\frac{g_ig_j}{\sqrt{M_iM_j}}\sqrt{\frac{(E_i+M_i)}{2M_i} \frac{(E_j+M_j)}{2M_j}} ,
    \end{align}
    where the $C_{ij}$ coefficients are given in Table~\ref{cij} of the Appendix.

    \item The last ingredient remaining to evaluate the spin-half $\mathcal{T}$-matrix is the lowest order amplitude  describing the transition between pseudoscalar-baryon and vector-baryon channels. We obtain from Eq.~\ref{pbvb}   
    \begin{align}
    \label{kroll}
        &\mathcal{V}_\text{PBVB}\left(i\to j\right)=i \sqrt{3}~g_{KR}~ \frac{A_{ij}}{2f_\pi^2}\sqrt{\frac{(E_i+M_i)}{2M_i} \frac{(E_j+M_j)}{2M_j}},
    \end{align}
where the coefficients $A_{ij}$ are listed in Table~\ref{aij} of the appendix.

\end{itemize}

Next, vector-baryon systems can have total spin 3/2 as well. We have studied S-wave interactions in this configuration too, and we get the following expressions for the lowest amplitudes corresponding to the $CT$-term and $u$-channel,
       \begin{align}
    \label{ct3}
      \mathcal{V}_{CT,\text{VB}}^{3/2}\left(i\to j\right)=-\frac{C_{ij}}{2}\frac{g_ig_j}{\sqrt{M_iM_j}}\sqrt{\frac{(E_i+M_i)}{2M_i} \frac{(E_j+M_j)}{2M_j}},
    \end{align}
    and
   \begin{align}
         \label{u3vb}
        \mathcal{V}_{u,\text{VB}}^{3/2}\left(i\to j\right)=U_{ij,\text{VB}}~ \frac{2g_i g_j}{2\tilde{M}-m_v}\sqrt{\frac{(E_i+M_i)}{2M_i} \frac{(E_j+M_j)}{2M_j}}.
    \end{align}
The values $C_{ij}$ and $U_{ij,\text{VB}}$ do not depend on the spin of the system, and are as given in Tables.~\ref{cij}, and~\ref{uij} of the Appendix. It might be useful to remind the reader here that the $t$-channel interaction is the same as given by Eq.~(\ref{eq:t}).

It remains to discuss the calculation of the loop function, $G$, needed to calculate Eq.~(\ref{bseq}), which is ultraviolet divergent and therefore requires regularization. We follow the dimensional regularization scheme, in which the divergence is absorbed into an unknown subtraction constant, and the corresponding expression is
\begin{align}\label{loop}
G_i (\sqrt{s}, m_i^2, M_i^2) &=\frac{2M_i}{16\pi^2} \Biggl\{ a_i (\mu) + \ln \frac{M_i^2}{\mu^2} + \frac{m_i^2-M_i^2+s}{2s}\ln \frac{m_i^2}{M_i^2} \Biggr.\\\nonumber
&+ \frac{\bar{q}_i}{\sqrt{s}} \Bigl[ \ln\left(s- \left( M_i^2 - m_i^2 \right) + 2\bar{q}_i\sqrt{s}\right) \Bigr.+  \ln\left(s+ \left( M_i^2 - m_i^2 \right) + 2\bar{q}_i\sqrt{s}\right) \\\nonumber
&- \ln\left(-s +\left( M_i^2 - m_i^2 \right) + 2\bar{q}_i\sqrt{s}\right)- \Biggl. \Bigl. \ln\left(-s- \left( M_i^2 - m_i^2 \right) + 2\bar{q}_i\sqrt{s}\right) \Bigr] \Biggr\},
\end{align}
where the regularization scale is set to be $\mu = 630 \text{ MeV}$. 
To estimate model uncertainties, we will show the results for two different sets of values of the subtraction constants, $a_i$, in the next section.

Finally, we must mention that the vector mesons in our system have a finite decay width, and we take it into account by  convoluting the two-body loop function over these widths. We do that by varying the mass of the mesons in question, following Ref.~\cite{Khemchandani:2018amu}, within a range allowed by twice their widths.

For the purpose of identification of states generated by the dynamics in the system, we search for poles in the complex energy plane.  We do this by continuing analytically the amplitudes to the complex plane. By recalling that the kernels are real, the procedure involves writing the loop function, above the threshold of the $i$th channel, as
\begin{align}
  G_i(\sqrt{s}-i\eta)= G_i(\sqrt{s}+i\eta)-i 2\text{Im\{}G_i(\sqrt{s}+i\eta)\}.
\end{align}

In the next section, we present the results for the modulus squared amplitudes obtained by solving Eq.~(\ref{bseq}. To encourage future discussions and experiments concerning the triple strangeness sector, we have also calculated the femtoscopic correlation function.

\section{Results and discussions}
We begin by showing the modulus squared amplitudes for different channels for real energies, which can be related to the corresponding cross sections. A peak structure in such plots can indicate the formation of a state, which can be better interpreted by investigating the nature of the corresponding pole appearing in the complex plane.
\subsection{Amplitudes}
\begin{figure}[h!]
        \centering
        \includegraphics[width=0.85\linewidth]{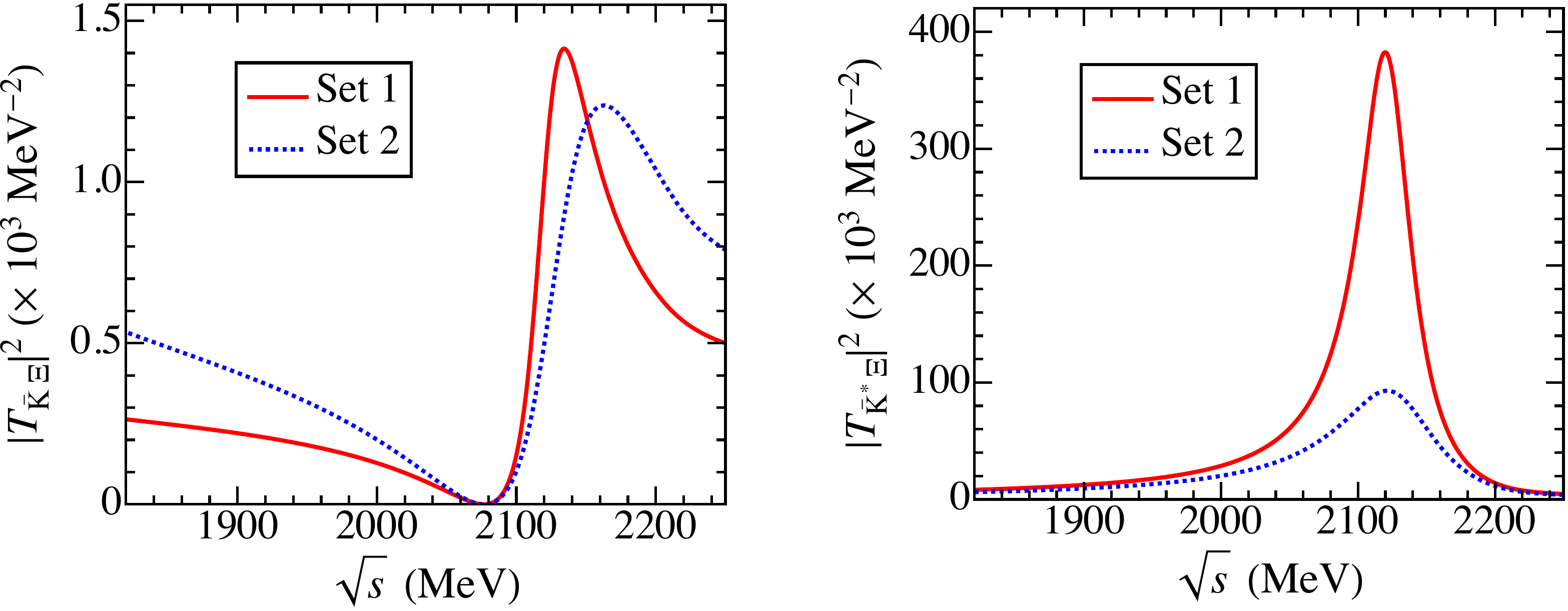}
        \caption{Modulus squared amplitudes projected on the isospin basis, with total $I = 0$ and spin half. The thick (red) line represent the results with the first set of values of the parameter $a_i$ (see Table~\ref{subs}) and the dotted (blue) line represent the second set. It should be noted that a factor 1000 has been multiplied to the squared amplitudes, as indicated in the labels at the vertical axes.}
        \label{AMP_SETS}
    \end{figure}

We present the results of the modulus squared amplitudes with total spin $J=\frac{1}{2}$ and total isospin, $I=0$ in Fig.~\ref{AMP_SETS}.
    We show the results for two sets of values of $a_i$ used in Eq.~(\ref{loop}). The values are  summarized in Table~\ref{subs}. 
       \begin{table}[h!]
        \centering
        \caption{Sets of values for the parameter $a_i$ in Eq.~(\ref{loop})}
        \begin{tabular}{c|c c}
        \hline \hline 
        & Set 1 & Set 2 \\
        \hline 
        $a_{\bar K \Xi}$ & $-3.1$ & $-2.0$\\
        $a_{\bar K^* \Xi}$ & $-2.0$ & $-2.0$ \\ \hline \hline
        \end{tabular}
        \label{subs}
    \end{table}
    It can be seen that Set 2 corresponds to the standard ``natural  values"~\cite{Ramos:2002xh,Oset:1997it,Hyodo:2008xr,Hyodo:2011qc,Khemchandani:2011et,Khemchandani:2011mf,Khemchandani:2013nma}.  As can be seen in Fig.~\ref{AMP_SETS}, a peak around $2.1 \text{ GeV}$ appears for both cases, Set 1 and Set 2, although with a different width. This shows that a state with mass around 2.1 GeV arises from the coupled channel dynamics, for different sets of the value of the regularization parameter. In particular, the values of Set 1 are chosen to fine tune the width of the state to get a better agreement with the properties of $\Omega(2109)$ found by the BESIII collaboration~\cite{BESIII:2024eqk}. It is useful to note that the subtraction constant, $a_i$, can be related to an equivalent three momentum cutoff, whose magnitude reflects the characteristic spatial scale of hadronic systems. Notably, the widely adopted choice $a=-2$, together with the regularization scale $\mu=630$~MeV, corresponds to a cutoff of in the range 600-700 MeV, depending on the channel. The value of $a=-3.1$, on the other hand, is related to a cutoff of the order of 2000 MeV.  The drift of the regularization parameter from its natural value can be interpreted~\cite{Hyodo:2008xr,Hyodo:2011qc,Ramos:2002xh} as the inclusion of a missing mechanism. In our case, the missing information could be a channel, like $\bar K\Xi(1530)$, whose $D$-wave scattering contribution has been considered in Ref.~\cite{Han:2025gkp}. The former work does not consider vector-baryon channels and finds a state with spin-parity $1/2^-$, and $M-i\Gamma/2=\left(2052\pm 20\right)-i\left(13\pm 2\right)$~MeV. One could also wonder if $\bar K^*\Xi^*$ could also be included in the coupled channel basis, which has been found to play an important role in the description $\Omega(2012)$~\cite{Li:2026wck,Lu:2022puv}. However, the threshold of $\bar K^*\Xi^*$ lies more than 300 MeV away from the nominal mass of $\Omega(2109)$~\cite{BESIII:2024eqk}. Hence, in the case of $\Omega(2109)$ including channels, such as $\bar K\Xi(1530)$ could be more useful. Investigating these possibilities should be carried out in the future.

\begin{figure}[h!]
     \centering
        \includegraphics[width=0.45\textwidth]{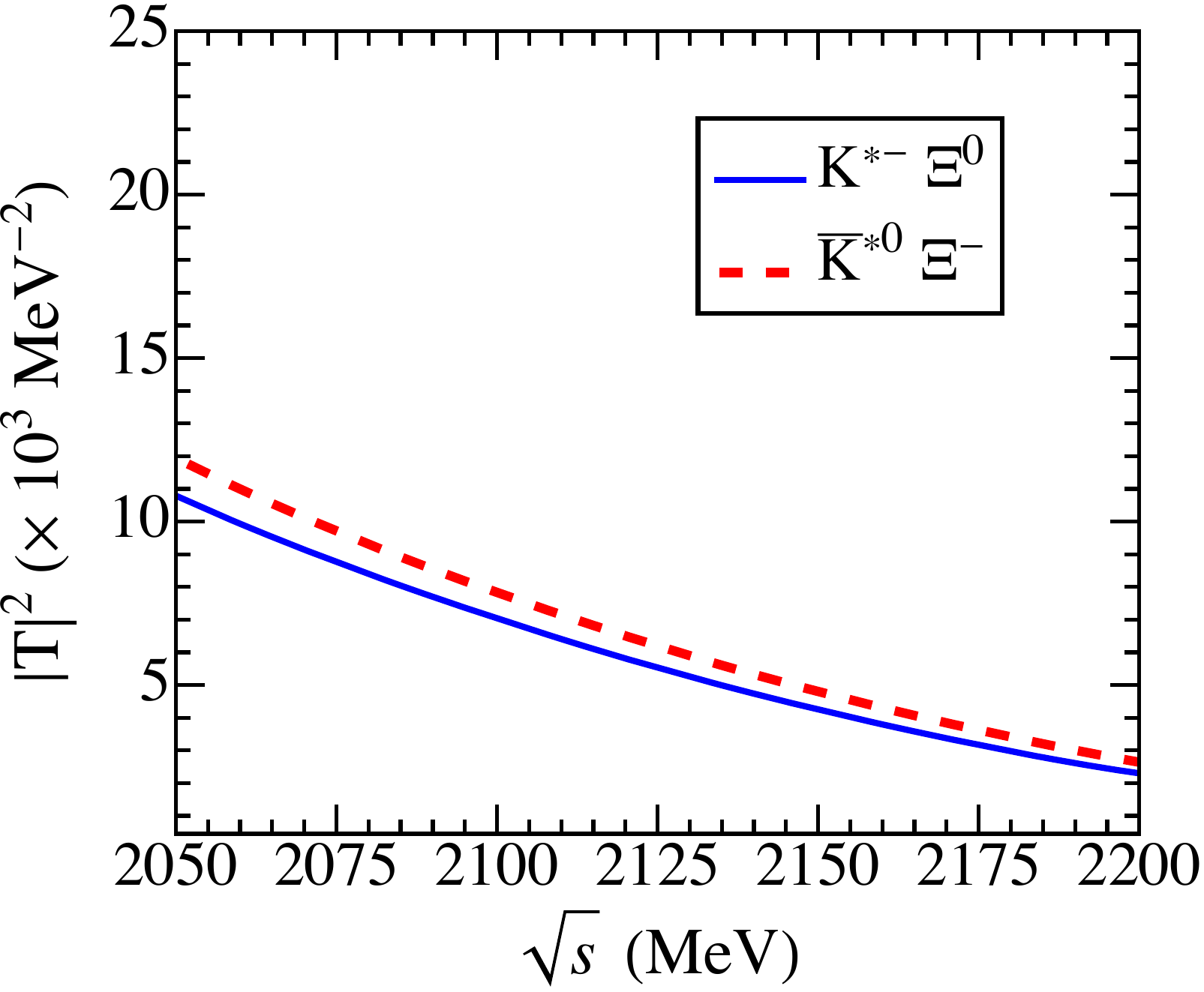}
        \caption{Modulus squared amplitudes in the particle basis for total spin $J=\frac{3}{2}$.}
        \label{AMP3}
    \end{figure}

Our findings coincide with those of Ref.~\cite{Su:2024lzy} also, where a $1/2^-$ state is found using QCD sum rules with mass $2.07\pm0.07$ GeV. The authors of the same work find a state with spin-parity $3/2^-$ as well, which could be related to $\Omega(2012)$. We do not find any state with spin 3/2 (see Fig.~\ref{AMP3}), since the interaction in this case is repulsive in our model. 
    It should be also added that no state is found in the isospin 1 configuration, if we use isospin average masses in the calculations. However, when we solve the Bethe-Salpeter equation in the particle basis, using different masses for each particle, we do find a structure from a small isospin violation.

\begin{figure}[h!]
        \centering
        \includegraphics[width=0.65\textwidth]{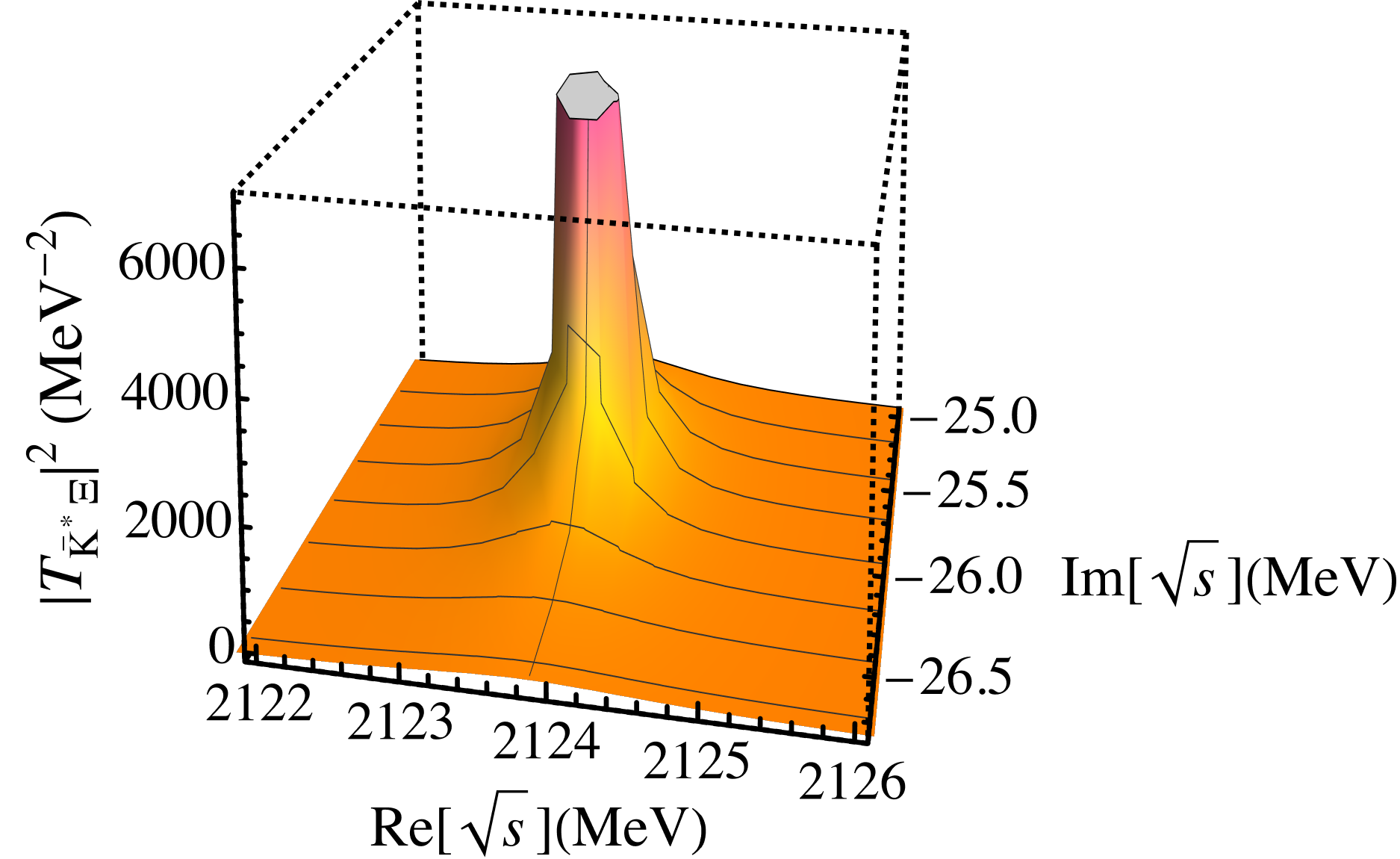}
        \caption{The $\bar K^* \Xi$ modulus squared amplitude, on the complex plane, in the isospin basis, with $I=0$. The pole position here is given by $\text{M}-i\frac{\Gamma}{2}=2123.75 \text{ MeV}-i25.75\text{ MeV}$.}
        \label{POLE}
    \end{figure}

  We end this section by showing the pole corresponding to the $J^P=1/2^-$amplitude obtained by using Set 1 in Fig.~\ref{POLE}. The pole position for this state is in good agreement with the newly reported excited state  $\Omega(2109)$~\cite{BESIII:2024eqk}. Although the experimental collaboration has assumed the new state to be a conventional $qqq$ state, such a state has not been found in quark models~\cite{Isgur:1978xj,Crede:2013kia,Xiao:2018pwe}. Our work shows that at least a large contribution to the dynamical origin of $\Omega(2109)$  must come from meson-baryon interactions. 
As can be already expected from the results shown in Fig.~\ref{AMP_SETS}, the resonance strongly couples to the $\bar K^*\Xi$ channel. We verified this by evaluating the residues of the pole (see Table~\ref{couplings}), which are interpreted as the coupling of the resonance to different channels.
\begin{table}[h!]
        \centering
        \caption{Residues of the four couple channels present in this study to the pole shown in Fig.~\ref{POLE}.}
        \begin{tabular}{p{1.5cm} |  p{2cm} p{2cm}}
        \hline \hline
         & $\text{Re}[\,g\,]$ & $\text{Im}[\,g\,]$  \\
         \hline 
        $K^-\Xi^0$ & $0.63$ & $-0.34$ \\
        $\bar K^0\Xi^-$ & $-0.61$ & $0.33$ \\
        $K^{*-}\Xi^0$ & $2.83$ & $0.21$ \\
        $\bar K^{*0}\Xi^-$ & $-2.87$ & $-0.19$ \\
        \hline \hline 
        \end{tabular}
        \label{couplings}
    \end{table}
These couplings can be used to calculate, for instance, the decay rate of the resonance ($R)$ to an allowed final state ($R\to 1+2$), as 
\begin{align}
    \Gamma_{R\to 1+2}=\frac{1}{2\pi}\frac{M_2}{M_R} P_{cm}\left|g\right|^2,
\end{align}
where the index 1(2) stands for a meson (baryon), $P_{cm}$ is the center of mass momentum of the decay channel and $g$ is the coupling of $R$ to the channel (see Table.~\ref{couplings}).

\subsection{Correlation Function}

    In recent times, the two particle femtoscopic correlation function (CF) has emerged, from both experimental collaborations and theoretical works, as a convenient new tool to investigate the properties of interactions among hadrons~\cite{Liu:2024uxn}.  
    
    In the essence, the CF can be understood as a measurement of attraction/repulsion in hadron-hadron systems~\cite{Liu:2024uxn,Agatao:2025ckp}. More precisely, the CF can be described as the ratio of the probability distribution of a pair of two hadrons measured in the same event (same collision), as a function of their relative momentum, and the probability of an uncorrelated pair coming from mixed events having the same relative momentum. 
    The interpretation of the CF results is, however, not always straightforward (see more details in Ref.~\cite{Liu:2024uxn}).

    To evaluate the correlation function, we make use of the Koonin-Pratt formula (Ref.~\cite{Vidana:2023olz,Feijoo:2023sfe,Albaladejo:2023pzq,Khemchandani:2023xup,Abreu:2024qqo,Liu:2025wwx}) generalized to include coupled channel interactions, 
  \begin{align}
        C_i(p_i)&=1+4\pi\theta(\Lambda-p_i)\int_0^{\Lambda}dr r^2 \mathcal{S}(r,R)  \nonumber\\
        &\times \biggl\{ -j_0^2(p_ir)+\sum_j |\mathcal{T}_{ij}(\sqrt{s})\tilde{G}_j(\Lambda,r,\sqrt{s})+j_0(p_ir)\delta_{ij}|^2\biggr\},\label{CFeq}
    \end{align}
    where $p_i $ is the relative momentum of the two particles, $j_0$ is the spherical Bessel function, $\mathcal{S}(r,R)$ is the source function with a Gaussian profile, and $R = 1 \text{ fm}$ is a source size parameter. The remaining ingredient in Eq.~\ref{CFeq} is the loop function $\tilde{G}_j(\Lambda,r,\sqrt{s})$

    \begin{eqnarray}
        \tilde{G}_j(\Lambda,r,\sqrt{s})= 2M\int\limits_0^{\Lambda}\frac{d^3q}{(2\pi)^3}\frac{E^B_j+E^M_j}{2E^B_jE^M_j}\frac{j_0(qr)}{s-(E^B_j+E^M_j)^2 + i\eta},
    \end{eqnarray}
where $E_j^{B(M)}=\sqrt{q^2+M_j(m_j)^2}$ is the energy of the baryon (meson) in channel $j$, and $\Lambda$ is a cut-off of the order of 1 GeV. 

\begin{figure}[h!]
        \centering        \includegraphics[width=\textwidth]{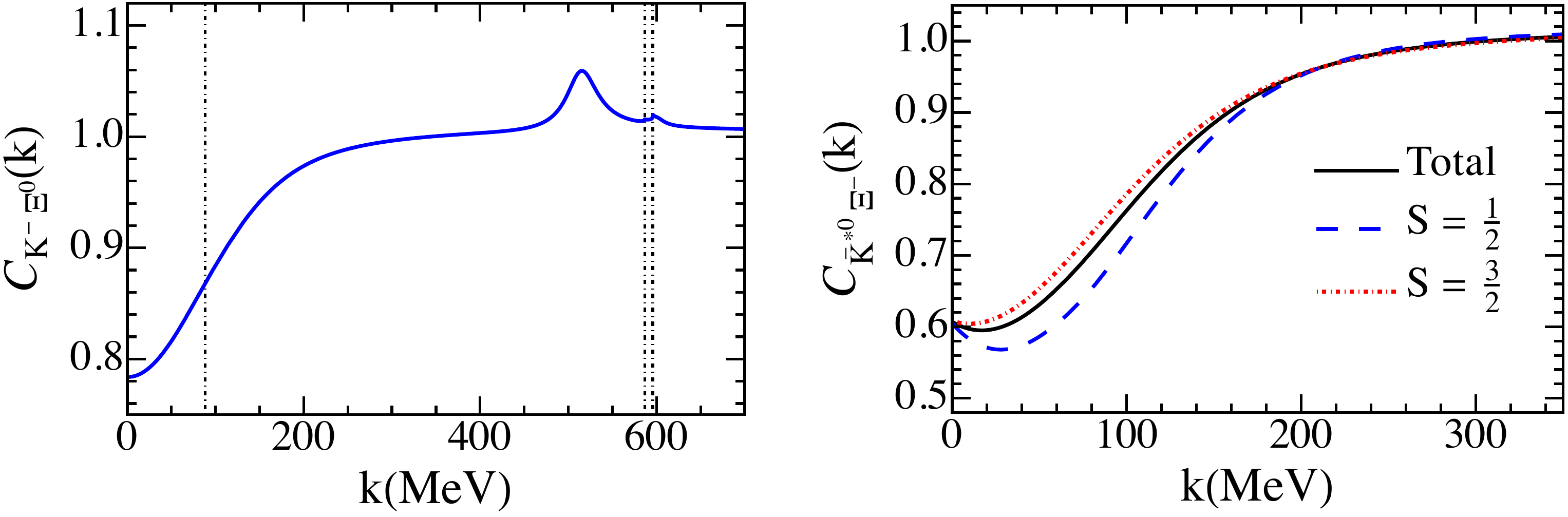}
        \caption{Correlation function for $K^-\Xi^0$ (left panel)  and $\bar K^{*0}\Xi^-$ (right panel).  }
        \label{CF1}
    \end{figure}

 We show the correlation function for the  $K^-\Xi^0$  and $\bar K^{*0}\Xi^-$ channels in Fig.~\ref{CF1}, as an example. The signal for $\Omega(2109)$ can be seen, in the form of a peak, around 500 MeV of c.m momentum in the correlation function of $K^-\Xi^0$. One can also notice a modification in the line shape of the CF due to the opening of the vector-baryon channels (having a small mass difference in the particle basis). The opening of these thresholds is marked by vertical lines in the left panel of Fig.~\ref{CF1}.

In case of the correlation function for the vector-baryon channel, we must remember that the system can have total spins 1/2 and 3/2 when interacting in the S-wave. Thus, in order to make useful predictions for future experimental investigations, a spin averaged observable must be calculated~\cite{Abreu:2024qqo}, as
\begin{align}
C_i(p_i)=\frac{1}{3}C_i^{S=1/2}(p_i)+\frac{2}{3}C_i^{S=3/2}(p_i).
\end{align}
We show the results for the case of total spin 1/2, 3/2, as well as the spin averaged correlation function in the right panel of Fig.~\ref{CF1}. In the spin $1/2$ case, the state lies below the VB threshold, as indicated by the shape of the dashed line in the right panel of Fig.~\ref{CF1} . Hence, the CF takes values below unity at low momenta. This behavior is expected  whenever a state below the threshold appears in the system whose correlation function is being studied~\cite{Liu:2025wwx,Liu:2024uxn}. One might wonder if the spin 3/2 CF, shown by a dotted line in the right panel of Fig.~\ref{CF1}, also corresponds to the existence of a state below the threshold. However, in this case, the interaction is repulsive, and the shape of the CF corresponding to a repulsive interaction is similar to that where a state is formed below the threshold~\cite{Liu:2024uxn}. 

Summarizing, our results show that a neat signal of $\Omega(2109)$ can be seen in experimental data on the correlation function for $\bar K \Xi$, as well as for $\bar K^*\Xi$. The same can be useful in establishing  its relation to meson-baryon dynamics. We hope that our work can encourage the measurement of these observables in the near future.

\begin{figure}[h!]
        \centering
        \includegraphics[width=\textwidth]{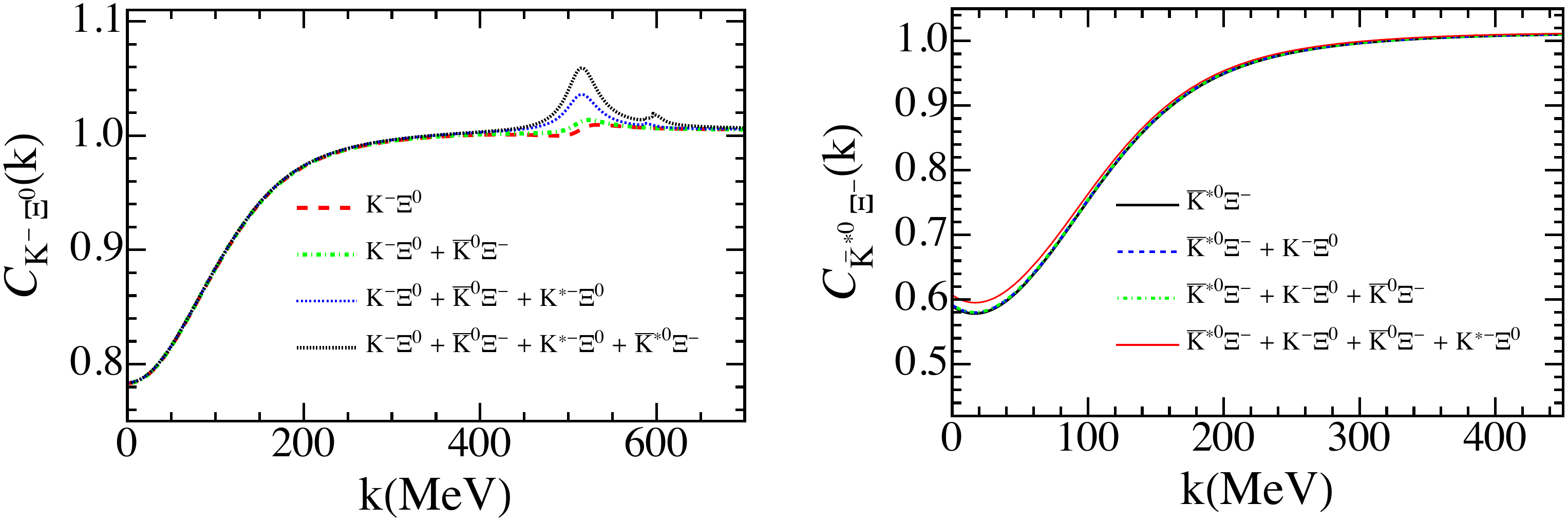}
        \caption{Contributions from different coupled channels to the correlation function. The right panel shows the spin averaged result for $\bar K^{*0}\Xi^-$.}
        \label{CF2}
    \end{figure}

We find it instructive  to show the contributions from the different channels to the correlation function in Fig.~\ref{CF2}. The useful information here is that the signal for $\Omega(2109)$ appears clearly in the correlation function of $K^-\Xi^0$ only when a transition to VB channels is added in Eq.~(\ref{CFeq}).

To end this manuscript, we also provide the scattering lengths of the $K^-\Xi^0$ and $\bar K^{*0}\Xi^-$ channels in Table~\ref{scatt}, which in the normalization we follow, is related to the $\mathcal{T}$-matrix as
\begin{align}
    a_i=-\frac{\mathcal{T}_{ii}(\sqrt{s_{thr}})}{4\pi}\frac{M_i}{\sqrt{s_{thr}}},
\end{align}
where $M_i$ is the mass of the baryon in the system. 
    \begin{table}[h!]
        \centering
        \caption{Scattering length values for the two types of channels considered in this work. The  total spin of the system is indicated as a subscript on $a$.}
        \begin{tabular}{p{1.8cm} | p{1.8cm}  p{1.8cm}|cc}
        \hline \hline 
        & $\text{Re}[\,a_{1/2}\,]$ & $\text{Im}[\,a_{1/2}\,]$&$\text{Re}[\,a_{3/2}\,]$ &$\text{Im}[\,a_{3/2}\,]$\\
        \hline
        $K^-\Xi^0$ & $-0.21 \text{ fm}$ & $0.00 \text{ fm}$&~~~--~~~&~~~~--~~~~\\
        $\bar{K}^{*0}\Xi^-$ & $-0.53 \text{ fm}$ & $0.25 \text{ fm}$&$-0.43$\text{ fm}&0.06\text{ fm}\\
        \hline \hline 
        \end{tabular}
        
        \label{scatt}
    \end{table}
In the convention we follow, a negative sign for the real part of the scattering length implies that the interaction is either repulsive or a state below the threshold is present in the system. Here, the signs of the $a^{\bar K\Xi}_{s=1/2}$ and $a^{\bar K^*\Xi}_{s=3/2}$ are related to a repulsive interaction, whereas that of $a^{\bar K^*\Xi}_{s=1/2}$ is due to the presence of the state related to $\Omega(2109)$ found below the threshold.
    
Information on the scattering length can be extracted from femtoscopic studies in heavy-ion collisions. The two-particle correlation function depends strongly on the scattering length and the effective range, and the values of these low energy scattering parameters are related to the underlying dynamics of the systems. We hope that the results obtained in this work will encourage new experimental studies in the near future.

\section{Acknowledgements}
This study was financed in part by the Coordena\c c\~ao de Aperfei\c coamento de Pessoal de N\'ivel Superior – Brasil (CAPES) – Finance Code 001.
The partial support from other Brazilian agencies is also gratefully acknowledged. We thank CNPq ( K.P.K: Grants No. 407437/ 2023-1 and No. 306461/2023-4; A.M.T: Grant No. 304510/2023-8), and CNPq/FAPERJ under the Project INCT-F\'{\i}sica Nuclear e Aplica\c c\~oes (Contract No. 408419/2024-5). H. N. is supported by the Grants-in-Aid for Scientific Research [Grant
No. 24K07051(C)]. A.H. is supported in part by Grants-in-Aid for Scientific Research [Grants No. 24K07050(C)].

\section{Appendix} 
In this appendix, we provide the coefficients appearing in the amplitudes used to solve the Bethe-Salpeter equation.
\begin{table}[h!]
    \centering
    \caption{$U_{ij,\text{VB}}$ coefficients of Eqs.~(\ref{uvb}) and~(\ref{u3vb}).}
    \begin{tabular}{c | c c}
    \hline \hline
    & $K^{*-}\Xi^0$ & $\bar{K}^{*0}\Xi^-$ \\ \hline
    $K^{*-}\Xi^0$ & $\frac{\big(2\tilde{M}+(D+F)m_v\big)^2}{8\tilde{M}^2}$ & $\frac{\big(-6\tilde{M}+(D-3F)m_v \big)^2}{48\tilde{M}^2}-\frac{\big(2\tilde{M}+(D+F)m_v\big)^2}{16\tilde{M}^2}$ \\

    $\bar{K}^{*0}\Xi^-$ & & $\frac{\big(2\tilde{M}+(D+F)m_v\big)^2}{8\tilde{M}^2}$ \\ \hline \hline
     
    \end{tabular}
    \label{uij}
\end{table}

\begin{table}[h!]
    \centering
    \caption{$U_{ij,\text{PB}}^{k}$ coefficients of Eqs.~(\ref{uchnpb}) and~(\ref{upb}). We write only the nonzero coefficients. }
    \begin{tabular}{c | p{3.2cm} p{3.2cm}}
    \hline  \hline
    & $K^-\Xi^0$ & $\bar{K}^0\Xi^-$  \\ \hline 
        
    $K^-\Xi^0$ & \begin{tabular}{c | c}
    $\Sigma^+$ & $2(D^\prime + F^\prime)^2$\\
    $\Sigma^-$ \\
    $\Sigma^0$ \\
    $\Lambda$ \\ \\
    \end{tabular} &

    \begin{tabular}{c | c}
    $\Sigma^+$ \\
    $\Sigma^-$ \\
    $\Sigma^0$ & $-(D^\prime+F^\prime)^2$ \\
    $\Lambda$ & $\frac{1}{3}(D^\prime-3F^\prime)^2$\\ \\
    \end{tabular} \\

    $\bar{K}^0\Xi^-$ & \begin{tabular}{c | c}
    $\Sigma^+$ \\
    $\Sigma^-$ \\
    $\Sigma^0$ & $-(D^\prime+F^\prime)^2$\\
    $\Lambda$ & $\frac{1}{3}(D^\prime-3F^\prime)^2$
    \end{tabular} &

    \begin{tabular}{c | c}
    $\Sigma^+$ \\
    $\Sigma^-$ & $2(D^\prime+F^\prime)^2$\\
    $\Sigma^0$ \\
    $\Lambda$ 
    \end{tabular} \\

    \hline \hline
     
    \end{tabular}
    \label{uijk}
\end{table}

\begin{table}[h!]
    \centering
    \caption{$C_{ij}$ coefficients of Eqs.~(\ref{ct}) and~(\ref{ct3}).}
    \begin{tabular}{c | c c}
    \hline \hline

    & $K^{*-}\Xi^0$ & $\bar{K}^{*0}\Xi^-$ \\ \hline

    $K^{*-}\Xi^0$ & $\frac{-(D+F)}{2}$ & $\frac{(D-F)}{2}$\\

    $\bar{K}^{*0}\Xi^-$ & & $\frac{-(D+F)}{2}$ \\ \hline \hline
    
    \end{tabular}
    \label{cij}
\end{table}

\begin{table}[h!]
    \centering
    \caption{$A_{ij}$ coefficients of Eq.~(\ref{kroll}).}
    \begin{tabular}{c | c c}
    \hline \hline

    & $K^{*-}\Xi^0$ & $\bar{K}^{*0}\Xi^-$ \\ \hline

    $K^{-}\Xi^0$ & $D^\prime+F^\prime$ & $-D^\prime+F^\prime$\\

    $\bar{K}^{0}\Xi^-$ & $-D^\prime+F^\prime$ & $D^\prime+F^\prime$ \\ \hline \hline
    
    \end{tabular}
    \label{aij}
\end{table}

\pagebreak
\bibliographystyle{elsarticle-num.bst}

\bibliography{RefsOMEGA}

\end{document}